\documentclass[12pt]{iopart}

\usepackage{iopams}

\newtheorem{lemma}{Lemma}
\newtheorem{theorem}[lemma]{Theorem}

\begin{document}
\title{Particle dynamics inside shocks in Hamilton--Jacobi equations}

\author{Kostya Khanin$^1$ and Andrei Sobolevski$^{2,3}$}
\address{$^1$ Department of Mathematics, University of Toronto,
  Toronto, Ontario, Canada}
\address{$^2$ Institute for Information Transmission Problems of the
  Russian Academy of Sciences, Moscow, Russia}
\address{$^3$ Laboratoire J.-V. Poncelet (UMI 2615 CNRS),
  Moscow, Russia}
\eads{\mailto{khanin@math.toronto.edu}, \mailto{sobolevski@iitp.ru}}

\begin{abstract}
  Characteristic curves of a Hamilton--Jacobi equation can be seen as
  action minimizing trajectories of fluid particles. %
  For nonsmooth ``viscosity'' solutions, which give rise to
  discontinuous velocity fields, this description is usually pursued
  only up to the moment when trajectories hit a shock and cease to
  minimize the Lagrangian action. %
  In this paper we show that for any convex Hamiltonian there exists a
  uniquely defined canonical global nonsmooth coalescing flow that
  extends particle trajectories and determines dynamics inside the
  shocks. %
  We also provide a variational description of the corresponding
  effective velocity field inside shocks, and discuss relation to the
  ``dissipative anomaly'' in the limit of vanishing viscosity.
\end{abstract}

\submitto{Philosophical Transactions of the Royal Society A}
\maketitle

\section{Introduction}
\label{sec:introduction}

The Hamilton--Jacobi equation
\begin{equation}
  \label{eq:1}
 \frac{\partial\phi}{\partial t}(t, \bi x)
  + H(t, \bi x, \nabla\phi(t, \bi x)) = 0
\end{equation}
plays an important role in a large variety of mathematical and
physical problems. %
Apart from analytical mechanics, it appears in description of a whole
range of extended dissipative systems featuring nonequilibrium
turbulent processes, from microscales of condensed matter and
statistical physics through mesoscale setting of free-boundary fluid
to macroscale cosmological evolution (see, e.g., a non-exhaustive
collection of references in~\cite{Bec.J:2007b}). %
The central issue in a study of nonlinear evolution for \eref{eq:1} is
to understand, both from the mathematical and physical points of view,
the behaviour of the system after the inevitable formation of
singularities. %

A theory of weak solutions for a general Hamilton--Jacobi equation,
employing the regularization by infinitesimal diffusion, exists since
the 1970s \cite{Kruzhkov.S:1975,Lions.P:1982,Crandall.M:1992}. %
In the one-dimensional setting this theory is essentially equivalent
to the earlier theory of hyperbolic conservation laws in fluid
mechanics, developed in the 1950s
\cite{Hopf.E:1950,Lax.P:1954,Oleuinik.O:1954}. %
In more than one dimension, however, the two theories are no longer
parallel. %

The theory of weak solutions for the Hamilton--Jacobi equation is
closely related to calculus of variations, and from this point of view
one can say that introduction of diffusion is motivated essentially by
stochastic control arguments \cite{Fleming.W:2005}. %
In the present paper we adopt a related but somewhat complementary
viewpoint, in which the Hamilton--Jacobi equation is considered as a
fluid dynamics model, and construct the flow of ``fluid particles''
inside the shock singularities of a weak solution. %

A useful example to be borne in mind when thinking about~\eref{eq:1}
--- and arguably the most widely known variant thereof --- is the
Riemann, or inviscid Burgers, equation
\begin{equation}
  \label{eq:2}
  \frac{\partial\bi u}{\partial t} + \bi u\cdot\nabla\bi u = 0,\qquad
  \nabla\times\bi u = 0.
\end{equation}
It is obtained for the Hamiltonian $H(t, x, \bi p) = |\bi p|^2/2$ by
setting $\bi u(t, \bi x) = \nabla\phi(t, \bi x)$. %
Note that in~\eref{eq:2} it is essential that the velocity field $\bi
u$ is curl-free, so this model is indeed equivalent to the Hamilton--Jacobi
equation \eref{eq:1}. %
The Riemann equation may in turn be considered as a limit of vanishing
viscosity of the Burgers equation
\begin{equation}
  \label{eq:3}
  \frac{\partial\bi u^\mu}{\partial t}
  + \bi u^\mu\cdot\nabla\bi u^\mu
  = \mu{\nabla}^2\bi u^\mu,\qquad  \nabla\times\bi u^\mu = 0,
\end{equation}
so solutions of~\eref{eq:2} can be defined as limits of smooth
solutions to~\eref{eq:3} as the positive parameter $\mu$ goes to
zero. %

The Burgers equation is in fact very special: it can be exactly mapped
by the Cole--Hopf transformation into the linear heat equation and therefore
explicitly integrated \cite{Hopf.E:1950}. %
Nonetheless the qualitative behaviour of solutions to a parabolic
regularization of~\eref{eq:1} for a general convex Hamiltonian
\begin{equation}
  \label{eq:4}
  \frac{\partial\phi^\mu}{\partial t} + H(t, \bi x, \nabla\phi^\mu)
  = \mu{\nabla}^2\phi^\mu
\end{equation}
in the limit of vanishing viscosity is similar to that for the Burgers
equation. %
It turns out that as $\mu \to 0$ there exists a limit $\phi(t, \bi x)=
\lim\phi^\mu(t, \bi x)$ which is called the \emph{viscosity
  solution}. %
Remarkably the viscosity solution can be described by a purely
variational construction which does not use the viscous regularization
at all. %
Below we briefly recall the main ideas of this variational approach. %

Assume that the Hamiltonian function $H(t, \bi x, \bi p)$ is smooth
and strictly convex in the momentum variable $\bi p$, i.e., is such
that for all $(t, \bi x)$ the graph of $H(t, \bi x, \bi p)$ as a
function of $\bi p$ lies above any tangent plane and contains no
straight segments. %
This implies that the velocity $\bi v = \nabla_{\bi p} H(t, \bi x, \bi
p)$ is a one-to-one function of~$\bi p$. %
In addition, the Lagrangian function
\begin{equation*}
%  \label{eq:5}
  L(t, \bi x, \bi v) = \max\nolimits_{\bi p}\,
  [\bi p\cdot\bi v - H(t, \bi x, \bi p)]
\end{equation*}
under the above hypotheses is smooth and strictly convex in~$\bi v$,
although it may not be everywhere finite: e.g., the relativistic
Hamiltonian $H(t, \bi x, \bi p) = \sqrt{1 + |\bi p|^2}$ corresponds to
the Lagrangian $L(t, \bi x, \bi v)$ that is defined for $|\bi v| \le
1$ as $-\sqrt{1 - |\bi v|^2}$ and should be considered as taking value
$+\infty$ elsewhere. %
This will not happen if in addition one assumes that the Hamiltonian
$H$ grows superlinearly in $|\bi p|$.%

The relation between the Lagrangian and the Hamiltonian is symmetric:
they are Legendre conjugate to one another. %
This relation can also be expressed in the form of the Young
inequality:
\begin{equation*}
%  \label{eq:6}
  L(t, \bi x, \bi v) + H(t, \bi x, \bi p) \ge \bi p\cdot\bi v.
\end{equation*}
This inequality holds for all $\bi v$ and~$\bi p$ and turns into
equality whenever $\bi v = \nabla_{\bi p}H(t, \bi x, \bi p)$ or $\bi p
= \nabla_{\bi v} L(t, \bi x, \bi v)$. %
The two functions $\nabla_{\bi p}H(t, \bi x, \bi p)$ and~$\nabla_{\bi
  v} L(t, \bi x, \bi v)$ are inverse to each other; we will call them
the \emph{Legendre transforms} at $(t, \bi x)$ of $\bi p$ and of~$\bi
v$. %
(Usually the term ``Legendre transform'' refers to the relation
between the conjugate functions $H$ and~$L$; here we follow the usage
adopted by A.~Fathi in his works on weak KAM theory
\cite{Fathi.A:2009}, which is more convenient in the present
context.) %

Note that if $H(t, \bi x, \bi p) = |\bi p|^2/2$, then $L(t, \bi x, \bi
v) = |\bi v|^2/2$ and the Legendre transform reduces to the identity
$\bi v \equiv \bi p$, blurring the distinction between velocities and
momenta. %
This is another very special feature of the Burgers equation. %

Now assume that $\phi(t, \bi x)$ is a strong solution of the inviscid
equation~\eref{eq:1}, i.e., a smooth function that satisfies
the equation in the classical sense. %
For an arbitrary trajectory $\bgamma(t)$ the full time derivative
of~$\phi$ along $\bgamma$ is given by
\begin{equation}
  \label{eq:7}
  \frac{\rmd\phi(t, \bgamma)}{\rmd t}
  = \frac{\partial\phi}{\partial t} + \nabla\phi\cdot \dot{\bgamma}
  = \nabla\phi\cdot \dot{\bgamma} - H(t, \bgamma, \nabla\phi)
  \le L(t, \bgamma, \dot{\bgamma}),
\end{equation}
where at the last step the Young inequality is used. %
This implies a bound for the mechanical action corresponding to the
trajectory $\bgamma$:
\begin{equation}
  \label{eq:9}
  \phi(t_2, \bgamma(t_2)) \le
  \phi(t_1, \bgamma(t_1)) +
  \int_{t_1}^{t_2}
  L(s, \bgamma(s), \dot{\bgamma}(s))\, \rmd s.
\end{equation}

Equality in~\eref{eq:7} is only achieved if $\dot{\bgamma}$ is the
Legendre transform of $\nabla\phi$ at $(t, \bgamma)$:
\begin{equation}
  \label{eq:8}
  \dot{\bgamma}(t) = \nabla_{\bi p}
  H(t, \bgamma, \nabla\phi(t, \bgamma)).
\end{equation}
This represents one of Hamilton's canonical equations, with momentum
given for the trajectory~$\bgamma$ by $\bi p_{\bgamma}(t) :=
\nabla\phi(t, \bgamma(t))$. %
The other canonical equation, $\dot{\bi p} = -\nabla_{\bi x} H$,
follows from \eref{eq:1} and~\eref{eq:8} because
\begin{equation*} \fl % for full left alignment
  \dot{\bi p}_{\bgamma}(t) = \frac{\partial\nabla\phi}{\partial t}
  + (\nabla\otimes\nabla\phi)\cdot\dot{\bgamma}
  = -\nabla_{\bi x}H(t, \bgamma, \nabla\phi)
  - \nabla_{\bi p}H\cdot (\nabla\otimes\nabla \phi)
  + (\nabla\otimes\nabla\phi)\cdot\dot{\bgamma}
\end{equation*}
Therefore the bound~\eref{eq:9} is achieved for trajectories
satisfying Hamilton's canonical equations. %
This is a manifestation of the variational \emph{principle of the
  least action}: Hamiltonian trajectories $(\bgamma(t), \bi
p_{\bgamma}(t))$ are (locally) action minimizing. %
In particular, if the initial condition
\begin{equation}
  \label{eq:10}
  \phi(t = 0, \bi y) = \phi_0(\bi y),
\end{equation}
is a fixed smooth function, the identity
\begin{equation*}
%  \label{eq:11}
  \phi(t, \bi x) = \phi_0(\bgamma(0))
  + \int_0^t L(s, \bgamma(s), \dot{\bgamma}(s))\, \rmd s
\end{equation*}
holds for a minimizer $\bgamma$ such that $\bgamma(t) = \bi x$. %

The least action principle can be used to
\emph{construct} the viscosity solution
corresponding to the initial data~\eref{eq:10}:
\begin{equation}
  \label{eq:12}
  \phi(t, \bi x) = \min\nolimits_{\bgamma\colon \bgamma(t) = \bi x}
  \Bigl(\phi_0(\bgamma(0))
  + \int_0^t L(s, \bgamma(s), \dot{\bgamma}(s))\, \rmd s\Bigr).
\end{equation}
This is the celebrated Lax--Oleinik formula (see, e.g.,
\cite{E.W:2000} or \cite{Fathi.A:2009}), which reduces a PDE problem
\eref{eq:1},~\eref{eq:10} to the variational problem~\eref{eq:12}
where minimization is extended to all smooth curves $\bgamma$ such
that $\bgamma(t) = \bi x$. %

The function $\phi$ defined by~\eref{eq:12} is smooth at those points
$(t, \bi x)$ where the minimizing trajectory is unique. %
In this case, the minimizer can be embedded in a smooth family of
minimizing trajectories whose endpoints at time $0$ and $t$ are
continuously distributed about $\bgamma(0)$ and $\bgamma(t) = \bi
x$. %
A piece of initial data $\phi_0$ gets continuously deformed according
to~\eref{eq:7} along this bundle of trajectories into a piece of
smooth solution $\phi$ to~\eref{eq:1} defined in a neighbourhood
of~$\bi x$ at time~$t$. %
Of course the Hamilton--Jacobi equation is satisfied in a strong sense
at all points where $\phi$ is differentiable. %

However, the crucial feature of~\eref{eq:12} is that generally there
will be points $(t, \bi x)$ with several minimizers $\bgamma_i$, which
start at different locations $\bgamma_i(0)$ but bring the same value
of action to~$\bi x = \bgamma_i(t)$. %
Just as above, each of these Hamiltonian trajectories will be
responsible for a separate smooth ``piece'' of solution. %
Thus for locations $\bi x'$ close to $\bi x$ the function $\phi$ will
be represented as a pointwise minimum of smooth pieces $\phi_i$:
\begin{equation*}
%  \label{eq:13}
  \phi(t, \bi x') = \min\nolimits_i \phi_i(t, \bi x').
\end{equation*}
As all $\bgamma_i$ have the same terminal value of action, all these
pieces intersect at $(t, \bi x)$: $\phi_1(t, \bi x) = \phi_2(t, \bi x)
= \dots = \phi(t, \bi x)$. %
Thus the neighbourhood of $\bi x$ at time~$t$ is partitioned into
domains where $\phi$ coincides with each of the smooth
functions~$\phi_i$ and satisfies the Hamilton--Jacobi
equation~\eref{eq:1} strongly. %
These domains are separated by surfaces of various dimensions where
two, or possibly three or more, pieces $\phi_i$ intersect and hence
$\phi$ is not differentiable. %
Such surfaces are called \emph{shocks}. %
Note that a function~$\phi$ defined by the Lax--Oleinik formula is
continuous everywhere, including the shocks; it is its gradient that
suffers a discontinuity. %
In general, there are infinitely many continuous functions that match
the initial condition~\eref{eq:10} and are differentiable and satisfy
the Hamilton--Jacobi equation~\eref{eq:1} apart from some shock
surfaces, just as~$\phi$.  %
A standard one-dimensional example of such nonuniqueness is provided
by $\phi_\alpha(t, x) = \min(\alpha|x| - \alpha^2 t/2, 0)$, which for
any $\alpha > 0$ satisfies the initial condition $\phi_\alpha(0, x) =
0$ and the equation $\partial\phi_\alpha/\partial t +
|\partial\phi_\alpha/\partial x|^2/2 = 0$ apart from the shock rays $x
= \pm\alpha t/2$, $x = 0$. %
What distinguishes the function~$\phi$ defined by~\eref{eq:12} from
all these ``weak solutions,'' and grants it with important physical
meaning, is that $\phi$ appears in the limit of vanishing viscosity
for the regularized equation~\eref{eq:4} with the initial
condition~\eref{eq:10} (see, e.g., \cite{Lions.P:1982}). %

For a smooth Hamiltonian it can be proved that once shocks are created
they never disappear, although they can merge with one another. %
Another important physical feature of viscosity solutions is that
minimizers can only merge with shocks but never leave them: all
minimizers coming to some $(t, \bi x)$ in~\eref{eq:12} originate at $t
= 0$. %
It is easy to see that this is not so in the above example of
$\phi_\alpha$, where minimizers emerge from $x = 0$ at all times $t >
0$. %

Moreover, in a solution~$\phi$ given by~\eref{eq:12} a minimizer that
has come to a shock cannot be continued any longer as a minimizing
trajectory. %
Indeed, wherever it comes, there will be other trajectories originated
at $t = 0$ that will bring smaller values of action to the same
location. %
Hence for the purpose of the least action description~\eref{eq:12}
Hamiltonian trajectories are discontinued as soon as they are absorbed
by shocks. %
The set of trajectories which survive as minimizers until time~$t > 0$
is decreasing with $t$, but at all times it is sufficiently large to
cover the whole space of final positions. %
% Those trajectories that have not merged with a shock by time~$t$
% form the flow at later times. %

Let us now adopt an alternative viewpoint and consider the
Hamilton--Jacobi equation as a fluid dynamics model, assuming that
Hamiltonian trajectories~\eref{eq:8} are described by material
``particles'' transported by the velocity field $\bi u(t,\bi x)$,
which is the Legendre transform of the momenta field $\bi p(t,\bi x)=
\nabla \phi(t,\bi x)$. %
From this new perspective it is no longer natural to accept that
particles annihilate once they reach a shock. %
Can therefore something be said about the dynamics of those particles
that got into the shock, notwithstanding the fact that their
trajectories cease to minimize the action? %
The problem here comes from the discontinuous nature of the velocity
field $\bi u$, which makes it impossible to construct classical
solutions to the transport equation $\dot{\bgamma}(t) = \bi u(t,
\bgamma)$. %

In dimension~$1$ the answer to the question above is readily
available. %
Shocks at each fixed $t$ are isolated points in the $\bi x$ space and
as soon as a trajectory merges with one of them, it continues to move
with the shock at all later times. %
This description is related to C.~Dafermos' theory of generalized
characteristics \cite{Dafermos.C:2005} which, in fact, can be extended
to a much more general situation of nonconvex Hamiltonians and systems
of conservation laws. %
However, in several space dimensions shocks become extended surfaces
and already for equation~\eref{eq:1} with a strictly convex
Hamiltonian dynamics of trajectories inside shocks is by no means
trivial. %
The main goal of the present paper is to describe a natural and
canonical construction of such dynamics. %

First results in this direction were obtained by I.~Bogaevsky
\cite{Bogaevsky.I:2004,Bogaevsky.I:2006} for the Burgers
equation~\eref{eq:3}. %
Bogaevsky suggested to consider the transport problem for a smooth velocity field $\bi u^\mu$:
\begin{equation*}
%  \label{eq:14}
  \dot{\bgamma}^\mu(t) = \bi u^\mu(t, \bgamma^\mu),\qquad
  \bgamma^\mu(0) = \bi y.
\end{equation*}
Since $\bi u^\mu$ for $\mu > 0$ is a smooth vector field, this
equation defines a family of particle trajectories forming a smooth
flow. %
The next step is to take the limit of this flow as $\mu\downarrow
0$. %
Bogaevsky proved that this limit exists as a non-differentiable
continuous flow, for which the forward derivative $\dot{\bgamma}(t +
0) = \lim_{\tau\downarrow 0}[\bgamma(t + \tau) - \bgamma(t)]/\tau$ is
defined everywhere. %
If $\bgamma(t)$ is located outside shocks, this derivative coincides
with $\bi u(t, \bgamma(t))$. %
Otherwise there are several limit values of velocity $\bi u_i$, and
Bogaevsky discovered an interesting explicit representation for
$\dot{\bgamma}(t + 0)$: it coincides with the center of the smallest
ball that contains all~$\bi u_i$. %
It should be remarked that uniqueness of a limit flow in the case of a
quadratic Hamiltonian was earlier observed by P.~Cannarsa and
C.~Sinestrari in the context of propagation of singularities for the
eikonal equation and differential inclusions
\cite[Lemma~5.6.2]{Cannarsa.P:2004}. %

The original approach in~\cite{Bogaevsky.I:2004,Bogaevsky.I:2006} is
based on the specific properties of the Burgers equation and cannot be
applied in the case of general convex Hamiltonians. %
In particular, the method uses the identity of velocities and momenta,
which does not hold in the general setting. %
At the same time the common wisdom says that all Hamilton--Jacobi
equations with convex Hamiltonians must have similar properties. %

In this work we consider the above strategy, consisting in the
parabolic regularization of the Hamilton--Jacobi equation and
investigation of the vanishing viscosity limit for the corresponding
regularized flow, in the case of general convex Hamiltonians. %
We show that such a limit exists, and derive an explicit
representation for the forward velocity of the limit flow that extends
the above result for the Burgers equation. %
Yet the mechanism powering these results in the general case is
completely different. %
It is based on the fundamental uniqueness of a possible limit
behaviour of~$\bgamma^\mu$, which we discuss in detail below.

The paper is organized as follows. %
In Section~\ref{sec:local-struct-visc} we study the local structure of
a viscosity solution near a singularity. %
We also introduce the notion of admissible velocity at a singularity
and show that it can be determined uniquely. %
Moreover, the unique admissible velocity provides a solution to a
certain convex minimization problem, which generalizes Bogaevsky's
construction of the center of the smallest ball. %
In Section~\ref{sec:vanish-visc-limit} we demonstrate that for any
nonsmooth viscosity solution there exists a unique continuous
nonsmooth flow of trajectories tangent to the field of admissible
velocities. %
Section~\ref{sec:concluding-remarks} contains concluding remarks. %

\ack

This work is supported by the NSERC Discover Grant and the Russian Foundation for Basic Research,
project RFBR-CNRS--07--01--92217; the second author is supported in
part by the French Agence Nationale de la Recherche, project
ANR--07--BLAN--0235 OTARIE. %
We acknowledge the hospitality of Observatoire de la C{\^o}te d'Azur
(Nice, France), where part of this work has been performed, as well as
support of the French Ministry for National Education. %

We are sincerely grateful to J{\'e}r{\'e}mie Bec, Patrick Bernard,
Ilya Bogaevsky, Yann Brenier, Philippe Choquard, and Boris Khesin for
numerous valuable discussions. %
It is our special pleasure to express gratitude to Uriel Frisch, who
made crucial contributions to the current revival of interest in the
``Burgers turbulence'' and related fields, and who is turning 70 in
2010. %

\section{Local structure of viscosity solutions and admissible
  momenta}
\label{sec:local-struct-visc}

Let $\phi$ be a viscosity solution to the Hamilton--Jacobi
equation~\eref{eq:1} with initial data~\eref{eq:10}. %
If there is a single minimizer coming to~$(t, \bi x)$, then $\phi$ is
differentiable at this point and
\begin{eqnarray*}
%  \label{eq:15}
  \phi(t + \tau, \bi x + \bxi)
  & = \phi(t, \bi x) + \frac{\partial\phi}{\partial t}\,\tau
  + \nabla\phi\cdot\bxi + \cdots \\
%  \label{eq:16}
  & = \phi(t, \bi x) - H(t, \bi x, \nabla\phi)\, \tau
  + \nabla\phi\cdot \bxi + \cdots,
\end{eqnarray*}
where dots $\cdots$ stand for higher-orger terms. %
If $\phi$ is not differentiable at $(t, \bi x)$, this means that there
are several minimizers $\bgamma_i$ such that $\bgamma_i(t) = \bi x$,
each bringing to~$(t, \bi x)$ a different piece $\phi_i$ of
solution. %
Then the Lax--Oleinik formula implies that
\begin{eqnarray*}
%  \label{eq:17}
  \phi(t + \tau, \bi x + \bxi)
  & = \min\nolimits_i \phi_i(t + \tau, \bi x + \bxi) \\
%  \label{eq:18}
  & = \phi(t, \bi x) + \min\nolimits_i (-H_i\tau + \bi p_i\cdot\bxi) + \cdots,
\end{eqnarray*}
where $\bi p_i := \nabla\phi_i(t, \bi x)$ and $H_i := H(t, \bi x, \bi
p_i)$. %

In the latter case neither of the expressions $-H_i \tau + \bi
p_i\cdot\bxi$ provides an adequate linear approximation to the
difference $\phi(t + \tau, \bi x + \bxi) - \phi(t, \bi x)$, but they
all \emph{majorize} this difference up to a remainder that is linear
or higher-order depending on $\tau$ and~$\bxi$. %
Evidently, so does too the linear form $-H\tau + \bi p\cdot\bxi$ for
any convex combination
\begin{equation*}
%  \label{eq:19}
  \bi p = \sum\nolimits_i \lambda_i \bi p_i,\qquad
  H = \sum\nolimits_i \lambda_i H_i
\end{equation*}
with $\lambda_i \ge 0$, $\sum_i \lambda_i = 1$. %
In convex analysis these convex combinations are called supergradients
of $\phi$ at $(t, \bi x)$ and the whole collecton of them, which is a
convex polytope with vertices $(-H_i, \bi p_i)$, is called the
\emph{superdifferential} of $\phi$
\cite{Cannarsa.P:2004,Rockafellar.R:1970}. %
We use R.~T.~Rockafellar's notation $\partial\phi(t, \bi x)$ for the
superdifferential \cite{Rockafellar.R:1970}. %

To avoid a possible misunderstanding it should be noted that, although
uniqueness of minimizer coming to~$(t, \bi x)$ implies
differentiability of~$\phi$ at $t$ and earlier times, it does
\emph{not} imply its differentiability at any $t + \tau > t$. %
The following example shows how this may happen. %
The function defined for $\tau \ge 0$ by
\begin{equation*}
%  \label{eq:20}
  \phi(t + \tau, x + \xi) = \phi(t, x)
  -\frac 43 \tau^3 - 2|\xi|\tau - \frac 43(|\xi| + \tau^2)^{3/2}
\end{equation*}
is a viscosity solution of the equation $\partial\phi/\partial t +
|\partial\phi/\partial x|^2/2 = 0$ that satisfies the smooth initial
condition $\phi(t, x + \xi) - \phi(t, x) = -\frac 43|\xi|^{3/2}$. %
For $\tau > 0$ a shock appears at $\xi = 0$, but differentiability at
$\tau = 0$ is recovered because $\partial\phi(t + \tau, x) =
\{-8\tau^2\}\times [-4\tau, 4\tau]$ shrinks to $(0, 0)$ as $\tau
\downarrow 0$. %
Such points $(t, \bi x)$ are called \emph{preshocks}
\cite{Bec.J:2007b}. %

The particular case of preshocks is an instance of a general fact:
replacing gradients with superdifferentials allows to recover
continuous differentiability, but in a weaker sense. %
Namely, suppose $(t_n, \bi x_n)$ converges to $(t, \bi x)$ and the
sequence $(-H_n, \bi p_n) \in \partial\phi(t_n, \bi x_n)$ has a limit
point $(-H, \bi p)$. %
By definition of superdifferential,
\begin{equation}
  \label{eq:21}
  \phi(t_n + \tau, \bi x_n + \bxi) - \phi(t_n, \bi x_n)
  \le -H_n\, \tau + \bi p_n\cdot\bxi + \cdots;
\end{equation}
passing here to the limit and using continuity of~$\phi$, we see that
$(-H, \bi p)\in \partial\phi(t, \bi x)$. %
Therefore the superdifferential $\partial\phi(t\, \bi x)$ contains all
the limit points of superdifferentials $\partial\phi(t_n, \bi x_n)$ as
$(t_n, \bi x_n)$ converges to $(t, \bi x)$. %

Suppose $(t, \bi x)$ is a point of shock where $k$ smooth branches
$\phi_i$ meet. %
It follows from the above discussion that for a particle moving from a
shock point $(t,\bi x)$ all possible values of the velocity $\bi v$
must be such that the corresponding momentum $\bi p(\bi v)$ belongs to
the convex hull of the available momenta $\bi p_i=\nabla \phi_i(t, \bi
x)$, $1\leq i \leq k$. %
However, one can say even more. %
For small positive $\tau$ not all the branches $\phi_i$ will
contribute to the solution $\phi$ at a point $(t+\tau, x+\bi v\tau)$,
but only those of them that correspond to a minimum in $\min_i(-H_i
+\bi p_i\cdot\bi v)$. %
Denote the corresponding set of indices
\begin{equation}
  \label{eq:24}
  I(\bi v) :=
  \{1\leq j\leq k\colon -H_j +\bi p_j\cdot\bi v
  = \min_i(-H_i +\bi p_i\cdot\bi v)\}.
\end{equation}
The set $I(\bi v)$ has the following physical meaning: if particle
moves away from a shock with a given velocity $\bi v$ then only
$\phi_j$ and $\bi p_j$ for $j\in I(\bi v)$ are relevant. %
Geometrically one can say that the convex hull of $\bi p_j$, $j \in
I(\bi v)$, is the $\bi p$-projection of the face of the
superdifferential $\partial\phi(t, \bi x)$ that looks toward an
infinitesimal observer who has just left~$(t, \bi x)$ with
velocity~$\bi v$. %

This implies that any possible velocity $\bi v$ must satisfy the
following condition.

\textbf{Admissibility Condition.} \emph{A velocity $\bi v^*$ is
  admissible if and only if the corresponding momentum belongs to the
  convex hull of the momenta $\bi p_j, j\in I({\bi v}^*)$.} %
Namely,
\begin{equation}
  \label{eq:27}
  \bi p^*(\bi v^*)=\sum_{j\in I(\bi v^*)} \lambda_j{\bi p}_j,
  \qquad \lambda_j\geq 0,
  \qquad \sum_{j\in I(\bi v^*)} \lambda_j = 1.
\end{equation}

To make the admissibility argument rigorous it is necessary to have
some control over the remainder term in~\eref{eq:21}. %
A natural function class that contains viscosity solutions of
Hamilton--Jacobi equations and in which such control is possible is
formed by semiconcave functions \cite{Cannarsa.P:2004}. %
We refer a reader interested in careful proofs of this and other
convex analytic results used in this paper to monographs
\cite{Cannarsa.P:2004,Rockafellar.R:1970}. %

Remarkably the Admissibility Condition determines the velocity $\bi
v^*$ uniquely.%

\begin{lemma}[Uniqueness]
  \label{lem:uniq}
  Let $\phi$ be a viscosity solution to the Cauchy problem
  \eref{eq:1},~\eref{eq:10}. %
  Then at any $(t, \bi x)$ there exists a unique admissible velocity
  $\bi v^*$. %
  Moreover this admissible velocity $\bi v^*$ is the unique point of
  the global minimum for the following function
  \begin{equation}
    \label{eq:29}
    \hat L(\bi v) := L(t, \bi x, \bi v) - \min\nolimits_i (-H_i + \bi
    p_i\cdot\bi v).
  \end{equation}
\end{lemma}

\noindent\textit{Proof.} %
Recall that $L(t, \bi x, \bi v)$ is a strictly convex function
of~$\bi v$ because of assumptions formulated in the Introduction. %
Rewriting
\begin{equation}
  \label{eq:30}
  L_i(\bi v) := L(t, \bi x, \bi v) + H_i - \bi p_i\cdot \bi v,\qquad
  \hat L(\bi v)
  = \max\nolimits_i\,L_i(\bi v),
\end{equation}
we see that $\hat L(\bi v)$ is a poitwise maximum of a finite number
of strictly convex functions and therefore is strictly convex
itself. %
Furthermore, because the Hamiltonian $H(t, \bi x, \bi p)$ is assumed
to be finite for all~$\bi p$, its conjugate Lagrangian $L(t, \bi x,
\bi v)$ grows faster than any linear function as $|\bi v|$ increases,
and all its level sets are bounded. %
Thus $\hat L(\bi v)$ attains its minimum at a unique value of
velocity~$\bi v^*$. %

Let us show that this point of minimum $\bi v^*$ satisfies the
admissibility condition. %
Indeed,
\begin{equation*}
  \nabla_{\bi v}L_i(\bi v^*)
  = \nabla_{\bi v}L(t, \bi x, \bi v^*) - \bi p_i
  = \bi p^*-\bi p_i.
\end{equation*}
Suppose that $\bi p^*$ does not belong to the convex hull of $\bi
p_j$, $j\in I(\bi v^*)$. %
Then there exists a vector $\bi h$ such that $(\bi p^* - \bi p_j)
\cdot \bi h < 0$ for all $j\in I(\bi v^*)$. %
It follows that $L_j(\bi v^* + \epsilon\bi h) < L_j(\bi v^*)$ for all
$j\in I(\bi v^*)$ if $\epsilon > 0$ is sufficiently small. %
Hence, $\hat L(\bi v^* + \epsilon\bi h) < \hat L(\bi v^*)$ for
sufficiently small $\epsilon$, which contradicts our assumption that
$\bi v^*$ is a point of minimum. %
This contradiction proves that $\bi v^*$ is admissible. %

To prove uniqueness we show that if $\hat \bi v$ is admissible then it
is a unique point of global minimum for the function $\hat L$. %
Using the strict convexity of $\hat L$, we obtain
\begin{eqnarray*}
  L_j(\hat{\bi v} + \bi h)
  & = L(t, \bi x, \hat{\bi v} +\bi h) + H_j
  - \bi p_j\cdot(\hat{\bi v} +\bi h) \\
  & > L_j(\hat{\bi v})
  + \nabla_{\bi v}L(t, \bi x, \hat{\bi v})\cdot\bi h
  - \bi p_j\cdot\bi h
  = L_j(\hat{\bi v}) + (\hat{\bi p} - \bi p_j)\cdot \bi h,
\end{eqnarray*}
where $\hat{\bi p}$ is the Legendre transform of $\hat{\bi v}$. %
Since $\hat{\bi v}$ is admissible, $\hat{\bi p} = \sum_j\lambda_j\bi
p_j$, where all $\lambda_j\geq 0$ and $\sum_j \lambda_j = 1$. %
Hence, $\sum_j\lambda_j(\hat{\bi p} - \bi p_j)\cdot \bi h =
[(\sum_j\lambda_j)\hat{\bi p} - \sum_j\lambda_j\bi p_j]\cdot h =
[\hat{\bi p} - \hat \bi p]\cdot h = 0$. %
It follows that $(\hat{\bi p} - \bi p_j)\cdot \bi h\geq 0$ for at
least one $j\in I(\hat \bi v)$. %
Thus $\hat L(\hat{\bi v} + \bi h) > \hat L(\hat{\bi v})$, which
implies that $\hat{\bi v}$ is a unique point of global minimum for
$\hat L$. %
This observation concludes the proof. %

The admissibility property, first formulated above in a somewhat
unmanageable combinatorial form~\eref{eq:27}, turns out to be the
optimality condition in a convex minimization problem given
by~\eref{eq:29}, i.e., a much simpler object. %
In particular, if $\phi$ is differentiable at $(t, \bi x)$, then $\hat
L(\bi v) = L(t, \bi x, \bi v) + H(t, \bi x, \nabla\phi) -
\nabla\phi\cdot\bi v$ and the minimum in~\eref{eq:29} is achieved at
the Legendre transform of~$\nabla\phi$. %
We thus recover Hamilton's equation~\eref{eq:8}. %

The following reformulation will clarify the connection between
admissibility and Bogaevsky's original construction for the Burgers
equation. %
Let $\bi v_i = \nabla_{\bi p} H(t, \bi x, \bi p_i)$ be the velocity
corresponding to the limit momentum~$\bi p_i$ and observe that $\bi
p_i = \nabla_{\bi v}L(t, \bi x, \bi v_i)$. %
The Young inequality implies that $H_i = H(t, \bi x, \bi p_i) = \bi
p_i\cdot \bi v_i - L(t, \bi x, \bi v_i)$ and therefore~\eref{eq:30}
assumes the form
\begin{equation*}
%  \label{eq:31}
  \hat L(\bi v) = \max\nolimits_i\,
  [L(t, \bi x, \bi v) - L(t, \bi x, \bi v_i)
  - \nabla_{\bi v}L(t, \bi x, \bi v_i)\cdot(\bi v - \bi v_i)].
\end{equation*}
The quantity in square brackets is known as the \emph{Bregman
  divergence} between vectors $\bi v$ and $\bi v_i$, a specific
measure of their separation with respect to the convex function~$L$
\cite{Bregman.L:1967}. %
When $L(t, \bi x, \bi v) = |\bi v|^2/2$, the Bregman divergence
reduces to (half) the squared distance between the two vectors; hence
the admissible velocity $\bi v^*$ exactly conicides with the centre of
smallest ball containing all~$\bi v_i$, and Bogaevsky's result is
recovered. %

Finally, we discuss the physical meaning of the function $\hat L$. %
Consider an infinitesimal movement from $(t, \bi x)$ with velocity
$\bi v$. %
Obviously $\phi(t, \bi x) + L(t, \bi x, \bi v)\, \rmd t - \phi(t +
\rmd t, \bi x + \bi v\,\rmd t) \geq 0$. %
It is easy to see that in the linear approximation in $\rmd t$
\begin{equation*}
  \phi(t, \bi x) + L(t, \bi x, \bi v)\, \rmd t
  - \phi(t + \rmd t, \bi x + \bi v\, \rmd t)
  = \hat L(\bi v)\, \rmd t.
\end{equation*}
Hence the unique admissible velocity $\bi v^*$ minimizes the rate of
the difference in action between the true minimizers and trajectories
of particles on shocks. %
In other words, the trajectory on a shock cannot be a minimizer but it
does its best to keep its surplus action growing as slowly as
possible.

\section{The vanishing viscosity limit}
\label{sec:vanish-visc-limit}

In the preceding section we constructed a canonical vector field $\bi
v^* = \nabla_{\bi p} H(t, \bi x, \bi p^*)$ corresponding to a given
viscosity solution $\phi$ of the Cauchy problem
\eref{eq:1},~\eref{eq:10}. %
The basis of this construction, the admissibility condition, appears
as a natural consistency condition
between velocities and supergradients. %
This condition together with the
variational principle~\eref{eq:29} guarantees uniqueness of
the admissible pair $(\bi v^*, \bi p^*)$. %

The vector field $\bi v^*(t, \bi x)$ can be decomposed into a union of
smooth tangent vector fields defined on connected pieces of smooth
shock surfaces of various dimensions as well as on the domain where
$\phi$ is differentiable, so dynamics in the latter domain or inside
any piece of a smooth shock surface is given locally by a smooth
flow. %
But globally the field $\bi v^*$ is discontinuous, and it is not
immediately clear if there exists an overall continuous flow of
trajectories $\bgamma$ that is compatible with~$\bi v^*$ in the sense
that $\dot{\bgamma}(t + 0) = \bi v^*(t, \bgamma)$. %
Even less obvious is the
uniqueness of such a flow. %

In order to answer these questions affirmatively we employ the
vanishing viscosity limit for the parabolic regularization
\begin{equation}
  \label{eq:32}
  \frac{\partial\phi^\mu}{\partial t} + H(t, \bi x, \nabla\phi^\mu)
  = \mu{\nabla}^2 \phi^\mu, \qquad \mu > 0,
\end{equation}
of the Hamilton--Jacobi equation~\eref{eq:1}. %
For sufficiently smooth initial data $\phi^\mu(t = 0, \bi y) =
\phi_0(\bi y)$ equation~\eref{eq:32} has a globally defined strong
solution, which is locally Lipschitz with a constant independent
of~$\mu$. %
Moreover, $\phi^\mu$ converges as $\mu \downarrow 0$ to the unique
viscosity solution~$\phi$ corresponding to the same initial data. %
Proof of these facts may be found, e.g., \cite{Lions.P:1982}, where
they are established for $\phi_0 \in C^{2, \alpha}$. %

Consider now the differential equation
\begin{equation}
  \label{eq:33}
  \dot{\bgamma}^\mu(t) = \nabla_{\bi p} H(t, \bgamma^\mu,
  \nabla\phi^\mu(t, \bgamma^\mu)), \qquad
  \bgamma^\mu(0) = \bi y.
\end{equation}
For $\mu > 0$ this equation has a unique solution $\bgamma^\mu_{\bi
  y}$, which continuously depends on the initial location~$\bi
y$. %
Fix a point $(t, \bi x)$ with $t > 0$ and pick for all sufficiently
small $\mu > 0$ trajectories $\bgamma^\mu$ such that $\bgamma^\mu(t) =
\bi x$. %
The uniform Lipschitz property of solutions~$\phi^\mu$ implies that
the curves $\bgamma^\mu$ are uniformly bounded and equicontinuous on
some interval containing~$t$. %
Hence there exists a curve $\bar{\bgamma}$ and a sequence $\mu_i
\downarrow 0$ such that $\lim_{\mu_i \downarrow 0} \bgamma^{\mu_i} =
\bar{\bgamma}$ uniformly. %
Note that all $\bgamma^{\mu_i}$ and~$\bar{\bgamma}$ are also Lipschitz
with a constant independent of~$\mu$ and that $\bar{\bgamma}(t) = \bi
x$. %
Let furthermore $\bar{\bi v}$ be a limit point of the ``forward
velocity'' of the curve~$\bar{\bgamma}$ at $(t, \bi x)$, i.e., let for
some sequence $\tau_k \downarrow 0$
\begin{equation*}
%  \label{eq:34}
  \bar{\bi v} = \lim\nolimits_{\tau_k\downarrow 0} \frac 1{\tau_k}
 [\bar{\bgamma}(t + \tau_k) - \bar{\bgamma}(t)].
\end{equation*}
%and let $\bar{\bi p}$ be the corresponding value of momentum:
%$\bar{\bi p} = \nabla_{\bi v} L(t, \bi x, \bar{\bi v})$. %

Of course neither the curve $\bar{\bgamma}$ nor the velocity $\bar{\bi
  v}$ are \textit{a priori} defined uniquely. %
Nevertheless it turns out that $\bar{\bi v}$ must satisfy the admissibility
condition
with respect to the solution~$\phi$
and therefore it coincides with $\bi v^*$. %
Also, trajectories of the flow $\bgamma^\mu$ converge as
$\mu\downarrow 0$ to segments of integral trajectories of the vector
field~$\bi v^*$ on smooth shock surfaces, establishing the uniqueness
of the limit flow~$\bar{\bgamma}$. %
Moreover, the limit flow is \emph{coalescing}: if two trajectories
intersect at time~$t$, they coincide for all $t' > t$. %
All these statements follow from the following fact. %

\begin{lemma}
  \label{lem:collapse}
  The flow defined by \eref{eq:33} for sufficiently small $\mu > 0$
  collapses the neighbourhood of the shock surface that passes through
  $(t, \bi x)$ and is tangent to~$(1, \bi v^*)$ onto this surface. %
\end{lemma}

Here is a sketch of proof. %
Let $\bi v^*$ be the admissible velocity corresponding to $(t, \bi
x)$, $\bi p^*$ the corresponding momentum, and define
\begin{equation}
  \label{eq:28}
  H^* = \bi
  p^*\cdot\bi v^* - \min_i (\bi p_i\cdot\bi v^* - H_i).
\end{equation}
The full time derivative of the function $\psi^\mu(t + \tau, \bi x +
\bxi) = \phi^\mu(t + \tau, \bi x + \bxi) - \bi p^*\cdot\bxi + H^*\tau$
along an integral trajectory $\bgamma^\mu$ of equation~\eref{eq:33}
which passes through $\bi x + \bxi$ at time $t + \tau$ is given by
\begin{equation*}
%  \label{eq:35}
  \frac{\rmd\psi^\mu(t + \tau, \bgamma^\mu(t + \tau))}{\rmd\tau}
  = \frac{\partial\phi^\mu}{\partial t} + H^*
  + (\nabla\phi^\mu - \bi p^*)\cdot \dot{\bgamma}^\mu.
\end{equation*}
% The function $\psi^\mu$ converges to $\psi$ as $\mu \downarrow 0$. %
Convergence of viscosity solutions implies convergence of their
superdifferentials (to see this, it is enough to replace
in~\eref{eq:21} the function $\phi$ with a sequence of functions
$\phi_n$ converging pointwise). %
Therefore limit points of $(\partial\phi^\mu/\partial t,
\nabla\phi^\mu)$ belong to~$\partial\phi(t, \bi x)$ for all $(t, \bi
x)$, and for sufficiently small $\mu > 0$, $\tau > 0$, $\bxi$ there
exists $(-H^\mu, \bi p^\mu)\in \partial\phi(t, \bi x)$ and $\bi v^\mu
= \nabla_{\bi p} H(t, \bi x, \bi p^\mu)$ such that, to the linear
approximation, $\bi v^\mu = \dot{\bgamma}^\mu + \cdots$ and
\begin{eqnarray}
  \label{eq:36}
  \frac{\rmd\psi^\mu}{\rmd\tau}
  & = -H^\mu + H^* + (\bi p^\mu - \bi p^*)\cdot \bi v^\mu + \cdots \\
  \label{eq:37}
  & = \bi p^\mu\cdot\bi v^* - H^\mu
  - \min_i (\bi p_i\cdot\bi v^* - H_i)\\
  \label{eq:38}
  &\qquad + (\bi p^\mu - \bi p^*)\cdot (\bi v^\mu - \bi v^*)
  + \cdots
\end{eqnarray}
Lines \eref{eq:37} and~\eref{eq:38} contain nonnegative quantities,
which are positive if $\bi p^\mu \neq \bi p^*$. %
To see this for~\eref{eq:38}, observe that by the strict convexity
$H(t, \bi x, \bi p^\mu) > H(t, \bi x, \bi p^*) + (\bi p^\mu - \bi
p^*)\cdot \bi v^*$ and $H(t, \bi x, \bi p^*) > H(t, \bi x, \bi p^\mu)
+ (\bi p^* - \bi p^\mu)\cdot \bi v^\mu$, and add these two
inequalities. %

On the other hand, the superdifferential of the function
\begin{equation*}
%  \label{eq:39}
  \lim_{\mu\downarrow 0}\psi^\mu(t + \tau, \bi x + \bxi)
  = \phi(t + \tau, \bi x + \bxi) - \bi p^*\cdot\bxi + H^*\tau
\end{equation*}
at $(\tau = 0, \bxi = 0)$ contains zero because $\bi p^*$ corresponds
to an admissible velocity. %
Therefore in the linear approximation the point $\bxi = 0$ and other
points of the shock surface of~$\psi$ tangent to $(1, \bi v^*)$ are
local maxima of~$\psi$ up to terms of higher order, and the full time
derivative of~$\psi$ along the curve $\bxi = \bi v^*\tau$ vanishes. %
As $\rmd\phi^\mu/\rmd t$ is positive for trajectories that start
outside the shock, we see that the flow~\eref{eq:33} collapses them
asymptotically on the shock surface. %
This completes the argument. %

The full details of this proof and the rigorous derivation of
uniqueness of the limit flow~$\bar{\bgamma}$ will be given in a
forthcoming article \cite{Khanin.K:2009}. %
Here we just formulate the main result. %

\begin{theorem}
  Let $\phi$ be a viscosity solution to the Cauchy problem for the
  Hamilton--Jacobi equation~\eref{eq:1} with initial
  data~\eref{eq:10}. %
  There exists a unique flow $\bgamma_{\bi y}$ of continuous
  trajectories such that $\bgamma_{\bi y}(0) = \bi y$,
  $\dot{\bgamma}_{\bi y}(t + 0)$ is defined for all $t$, $\bi y$ and
  coincides with the admissible velocity $\bi v^*(t, \bgamma_{\bi
    y}(t))$ given by solution to the convex minimization problem
 for~\eref{eq:29}. %
  The trajectory $\bgamma_{\bi y}$ continuously depends on~$\bi y$. %
  After a trajectory $\bgamma_{\bi y}$ comes to a shock, it stays
  inside the shock manifold for all later times. %
  The flow is coalescing: if two trajectories $\bgamma_{\bi y'}$ and
  $\bgamma_{\bi y''}$ coincide at time~$t$, then $\bgamma_{\bi y'}(t')
  = \bgamma_{\bi y''}(t')$ for all $t' > t$. %
\end{theorem}

% \section{The weak noise limit}
% \label{sec:weak-noise-limit}

\section{Concluding remarks}
\label{sec:concluding-remarks}

Starting from a viscosity solution $\phi$ to the Hamilton--Jacobi
equation, we have constructed a unique continuous coalescing flow
$\bgamma_{\bi y}$ compatible with the admissible velocity field $\bi
v^*$ defined in Section~\ref{sec:local-struct-visc} in the sense that
$\dot{\bgamma}_{\bi y}(t + 0) = \bi v^*(t, \bgamma_{\bi y})$. %
This flow is a natural extension of the smooth flow defined by
Hamilton's equation~\eref{eq:8}. %
We conclude with a few observations concerning this construction. %

1. Recall an important observation made in the original work of
Bogaevsky: pieces of the shock manifold, irrespective of their
dimension, are classified into \emph{restraining} and
\emph{nonrestraining} depending on whether $\bi p^*$ belongs to the
interior or the boundary of the convex polytope formed by projection
of~$\partial\phi(t, \bi x)$ on the $\bi p$ space. %
Particles stay on restraining shocks but leave non-restraining shocks
along pieces of shock manifold of lower codimension corresponding to
faces of the boundary containing~$\bi p^*$. %
Shocks of codimension one are always restraining; in particular, so
are all shocks in the one-dimensional case, which makes the
construction of the coalescing flow~$\bgamma_{\bi y}$ trivial, as
remarked in the introduction. %
Interestingly, this classification of shocks, introduced
in~\cite{Bogaevsky.I:2004} (``acute'' and ``obtuse''
superdifferentials of~$\phi$), seems to have been overlooked by
physicists before. %

2. Note that the construction of the admissible velocity $\bi v^*$
is purely kinetic: when the Lagragian is ``natural,'' i.e., has the
form $L(t, \bi x, \bi v) = K(\bi v) - U(t, \bi x)$, the value $\bi
v^*$ is the same for all choices of the potential term $U(t, \bi x)$
as long as kinetic energy~$K(\bi v)$ is fixed. %
We owe this observation to P.~Choquard.

3. Seen as a family of continuous maps of variational origin from
initial coordinates $\bi y$ to current coordinates $\bi x$, the
flow~$\bgamma_{\bi y}$ is clearly relevant for optimal transportation
problems \cite{Gangbo.W:1996,Villani.C:2009}. %
An interesting problem suggested by B.~Khesin is to study the extremal
properties of this flow. %
Indeed it is known from \cite{Khesin.B:2007} that before the first
shock formation the flow $\bgamma_{\bi y}$ is and action minimizing
flow of diffeomorphisms, while the first shock formation time $t^*$
marks a conjugate point in the corresponding variational problem. %
According to the suggested view, the flow constructed above may be
seen as a kind of saddle-point, rather than minimum, for a suitable
transport optimization problem. %

4. Another natural context to place our construction in is that of
differential inclusions (see, e.g., \cite{Aubin.J:1984}). %
The flow consructed here may be seen as a solution of differential
inclusion
\begin{equation}
  \label{eq:45}
  \dot{\bgamma} \in \nabla_{\bi p} H(t, \bgamma,
  \mathop{\mathrm{Pr}}\nolimits_{\bi p} \partial\phi(t, \bgamma)),
\end{equation}
where $\mathop{\mathrm{Pr}}_{\bi p}$ is the $\bi p$ projection of the
superdifferential~$\partial\phi$. %
In comparison with standard constructions of the theory of
differential inclusions the flow $\bgamma_{\bi y}$
solves~\eref{eq:45} in a stronger sense: the forward derivative
$\dot{\bgamma}_{\bi y}(t + 0)$ exists everywhere. %
Also, a simple modification of Lemma~\ref{lem:collapse} gives a proof
of uniqueness for inclusion~\eref{eq:45}. %

5. The flow $\bgamma_{\bi y}$ was constructed as a limit of a
parabolic regularization, and it was noticed above that the limit of
$(-\partial\phi^\mu/\partial t, \nabla\phi^\mu)$ belongs to
$\partial\phi(t, \bi x)$ for any~$(t, \bi x)$. %
This statement can be refined if one considers the values of
$\nabla\phi^\mu$ along trajectories of the flow~\eref{eq:33}. %
Namely, arguments of Section~\ref{sec:vanish-visc-limit} imply that
under the successive limits $\mu \downarrow 0$ and $\tau \downarrow
0$, the gradient $(\partial\phi^\mu/\partial t, \nabla\phi^\mu)$ taken
at $(t + \tau, \bgamma^\mu(t + \tau))$ converges to $(-H^*, \bi p^*)$.
Therefore
\begin{eqnarray*} \fl
%  \label{eq:40}
  \lim_{\tau\downarrow 0} \lim_{\mu\downarrow 0} \mu{\nabla}^2\phi^\mu
 (t + \tau, \bgamma^\mu(t + \tau))
  & = \lim_{\tau\downarrow 0} \lim_{\mu\downarrow 0}
  \Bigl[\frac{\partial\phi^\mu}{\partial t} + H(t, \bgamma^\mu(t + \tau),
  \nabla\phi^\mu)\Bigr] \\
%  \label{eq:41}
  & = H(t, \bi x, \bi p^*) - H^* \\
%  \label{eq:42}
  & = \min_i
  (\bi p_i\cdot\bi v^* - H_i) - L(t, \bi x, \bi v^*) \\
%  \label{eq:43}
  & = \hat L(\bi v^*) = \min\nolimits_{\bi v} \hat L(\bi v),
\end{eqnarray*}
where we took into account formulas~\eref{eq:28} and~\eref{eq:29}. %
In other words, minimum of the convex minimization
problem~\eref{eq:29} coincides with the value of the ``dissipative
anomaly'' in the parabolic regularization~\eref{eq:4} of the
Hamilton--Jacobi equation~\eref{eq:1}. %

6. Observe also that convergence of superdifferentials makes it
possible to use other smoothing procedures for $\phi$ (e.g.,
convoluting it with a standard mollifier), giving the same limit
$\bgamma_{\bi y}$. %
However, one can imagine the following completely different
regularization of the discontinuous velocity field $\nabla_{\bi p}
H(t, \bi x, \nabla\phi(t, \bi x))$. %
Physically speaking, this regularization may be characterized by a
``zero Prandtl number'' in contrast with the previous class of
regularizations featuring an ``infinite Prandtl number.'' %

Consider the stochastic equation
\begin{equation*}
%  \label{eq:44}
  \rmd\bgamma^\epsilon
  = \nabla_{\bi p} H(t, \bgamma^\epsilon,
  \nabla\phi(t, \bgamma^\epsilon))\, \rmd t + \epsilon\,\rmd\bi W(t),
\end{equation*}
where $\bi W$ is the standard Wiener process. %
The corresponding stochastic flow is well defined in spite of the fact
that $\nabla\phi$ does not exist everywhere: whenever the trajectory
$\bgamma^\epsilon$ hits shocks, noise in the second term will
instantaneously steer it in a random direction away from them. %

Assume that as $\epsilon\downarrow 0$ the stochastic flow
$\bgamma^\epsilon_{\bi y}$ tends to a limit flow $\tilde{\bgamma}_{\bi
  y}$, which is also forward differentiable. %
It is easy to see that, due to the averaging, the forward velocity
$\bi v^\dagger(t, \tilde{\bgamma}) := \dot{\tilde{\bgamma}}_{\bi y}(t
+ 0)$ must belong to the convex hull of $\bi v_j, j \in I(\bi
v^\dagger)$.  Namely,
\begin{equation}
  \label{eq:46}
  \bi v^\dagger = \sum_{j\in I(\bi v^\dagger)} \pi_j \bi v_j,
  \qquad \pi_j\geq 0,
  \qquad \sum_{j\in I(\bi v^\dagger)} \pi_j=1,
\end{equation}
where the velocities $\bi v_j(t, \bi x)$ are Legendre transforms of
the corresponding momenta $\bi p_j =\nabla\phi_i(t, \bi x)$ at a
singular point $(t, \bi x)$. %
The coefficients $\pi_i$ are equal to asymptotic values of shares of
time that a trajectory $\bgamma^\epsilon$ spends in each of the
domains where $\phi = \phi_j$. %
Condition~\eref{eq:46} above is another compatibility condition, in
certain sense dual to the admissibility condition in
Section~\ref{sec:local-struct-visc}. %
For want of a better term let us call such a velocity $\bi v^\dagger$
\emph{self-consistent}. %

The self-consistent velocity is a convex combination of
\emph{velocities} seen by an infinitesimal observer leaving $(t, \bi
x)$ with velocity~$\bi v^\dagger$. %
Compare this with the definition of admissible momentum $\bi p^*$,
which is a convex combination of \emph{momenta} see by a similar
observer moving vith velocity~$\bi v^*$. %
When $H(t, \bi x, \bi p) = |\bi p|^2/2$ and $\bi v = \bi p$,
self-consistent velocities and admissible velocities coincide. %
It is however clear that in the case of a general nonlinear Legendre
transform $\bi v^\dagger \neq \bi v^* = \nabla_{\bi p} H(t, \bi x, \bi
p^*)$. %

In view of the analogy between self-consistent velocities and
admissible momenta, it is tempting to conjecture that the admissible
velocity is also unique. %
Generally speaking, this statement is wrong, although it holds in one
and two space dimensions. %
In higher dimensions there exist Hamiltonians and sets of limiting
momenta for which there is more than one admissible velocity
\cite{Khanin.K:2009}. %
It is an interesting problem nevertheless to see whether a limiting
flow still exists in the limit of weak noise in spite of nonuniqueness
of admissible velocity. %
This problem carries a certain similarity with the problem of limit
behaviour for one-dimensional Gibbs
measures in the zero-temperature limit, in the case of nonunique ground states.

% Two kinds of singular points ("acute" and "obtuse")

% Purely kinetic effect (Choquard)

\end{document}